# TUM

INSTITUT FÜR INFORMATIK

Enhancing the SYSLAB System Model with State

Radu Grosu
Cornel Klein
Bernhard Rumpe

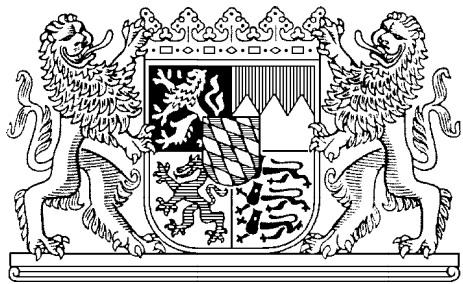

```
TUM-I9631
Juli 1996
```

TECHNISCHE UNIVERSITÄT MÜNCHEN

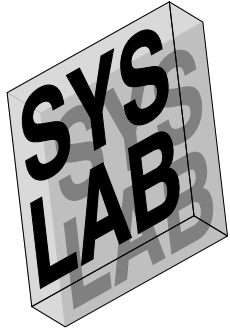

# Enhancing the SysLab System Model with State[1]

Radu Grosu, Cornel Klein, Bernhard Rumpe
Institut für Informatik, Technische Universität München
80333 München, Germany
email: (grosu|klein|rumpe)@informatik.tu-muenchen.de

TUM-I9631

[1]This paper originated in the SysLab project, which is supported by the DFG under the Leibnizpreis and by Siemens-Nixdorf

**Abstract**

In this report, the SysLab system model is complemented in different ways: State-box models are provided through timed port automata, for which an operational and a corresponding denotational semantics are given. Composition is defined for components modeled in the state-box view as well as for components modeled in the black-box view. This composition is well-defined for networks of infinitely many components. To show the applicability of the model, several examples are given.

# Contents

# Chapter 1

# Introduction

The SYSLAB project aims at providing a scientifically well-founded approach for software- and system development. Within SYSLAB, the system model (see [RKB95], [KRB96]) serves as a common reference model for the definition of the semantics of the SYSLAB description techniques, for defining notions of correctness for transformation rules and for the tool system to be developed.

The main emphasis of the technical report [RKB95] was to motivate the need for such a system model and to provide a common understanding and terminology for the project participants. Therefore, a very concise and abstract formal description of the system model has been given, while leaving intentionally many technical questions open. In particular, no glass-box view for elementary components has been given.

This report complements [RKB95] in many different ways. State-box models are provided by timed port automata, for which an operational and a corresponding denotational semantics are given. Composition is defined for components modeled in the state-box view as well as for components modeled in the black-box view. This composition is also well-defined for networks of infinitely many components. Also, some minor revisions concerning the system-model have been made. To show the applicability of the model, several examples are given.

The paper is organized as follows: In the following section, we repeat the main motivation for the SYSLAB system model from [RKB95]. The third section introduces streams as a basic mathematical concept for the paper. The fourth section is dedicated to basic components, i.e. to components that are not distributed. Here two views are presented, called the state-box view and the black-box view. In Section 5, networks of components are introduced. These allow to model the spatial or logical distribution of a system. In Section 6, the introduced techniques are used to specify the SYSLAB system model, which can be seen as a special system architecture appropriate for the kind of systems studied in SYSLAB. In Section 7, the applicability of the system model is demonstrated by two examples. The paper ends with a conclusion.

# Chapter 2

# Goals of the SysLab system model

Before we formally define the SysLab system model, we want to present the context in which the system model is used. Therefore, in the following section we briefly outline the state-of-the art in the area of software engineering in theory and practice, and give an overview over the goals of the SysLab project. In the remainder of this section, we put up requirements for the SysLab system model.

## 2.1 Software Engineering Methods

Methods for systems and software development, like OMT [RBP+92], Fusion [CAB+94], and GRAPES [Hel90], model a system in different abstraction levels and under different views. Within the process of modeling they use description techniques like entity-/relationship-diagrams and their object-oriented extensions, state automata, sequence charts or data-flow diagrams. A critical point of existing commercial methods is that the definition of the description techniques as well as the relationships between different description techniques of a method is usually only given informally. A lot of problems during the application of the methods exist, which are caused by the ambiguous and vague interpretation of the semantics of the used modeling concepts:

- the communication between the persons involved in the project is more difficult, because of ambiguities arising from informal semantic descriptions,
- it is impossible to define formal relationships between different description levels and to define rules to transfer information between two description levels,
- a solid basis for tool support is missing,

- and moreover even in one description level there is a lack of clarity concerning the consistency and completeness of a set of documents. Issues concerning "consistency" and "completeness" can only be tackled informally.

As a consequence, it is understandable why tool systems for methods ("CASE-Tools") often do not cause the expected gain in productivity: The information which can be acquired by the use of methods is, because of the deficient semantic foundation of the methods, not very evident. As a result of this the functionality of tools is mostly purely restricted to document editing and managing functions.

Recently, various approaches for formalizing methods of systems and software development were given. Well known are the so-called "meta-models", originating in the context of tool integration, (see [CDI92], [Tho89] and [HL93]). However, by this "models" almost only the abstract syntax of the description techniques is captured. An overview of several projects concerning the integration of structured methods with techniques of formal specification can be found in [SFD92]. In [Hus94], the British standard method SSADM [AG90] is formalized using the algebraic specification language SPECTRUM [BFG$^+$93]. The work of Hussmann goes further than the approaches described in [SFD92], because Hussmann states a mathematical model of the information systems modeled by SSADM to which he relates the different description techniques which occur in the method. This approach offers a complete analysis of the semantics of the SSADM-description techniques and their relationships, the definition of conditions for consistency and completeness of a set of description techniques, and a simple basis for obtaining prototypes by functional programs.

## 2.2 The role of the system model in SysLab

The SysLab-project aims at developing a practicable method for system- and software development. The method is required to be scientifically founded, such that it does not show the above-mentioned disadvantages due to the lack of a semantic foundation. Moreover, in SysLab a prototype a tool system will be created. The formalization should not end in itself, but it should provide the semantic basis for the check for consistency of the concepts. The semantic foundation is achieved by the usage of a uniform system model for SysLab. This abstract mathematical model of information processing systems is used as semantic basis for all description techniques defined in SysLab, such as object diagrams, state diagrams, data-flow diagrams, etc. The correctness of all transformation rules for the transformation of documents is defined in terms of the system model. Each document, such as an object diagram, is regarded as a proposition over the mathematical system model.

The formalization of description techniques leads primarily to a deeper comprehension of the meaning of the descriptions, the aspects on which statements are given, and their inter-relations. Therefore, description techniques could be used more objectively. Furthermore, it is possible to state conditions for consistency and completeness of a set of description documents, and to define and to analyze relationships between description documents of different abstraction levels. This is demonstrated in [Rum96]. Moreover, formalization is an important mile-stone on the way to a more effective tool support of methods, because semantic-preserving transformations between different description techniques are feasible which finally result in executable code. Finally, a flexible application of formal techniques, which is necessary in safety-critical applications, is possible.

## 2.3 Requirements on the system model

It is the aim of this paper to provide a common basis for all people involved in the SysLab-project concerning the notion of system used and concerning the definition of the semantic of the various description techniques. Therefore, the system model has to cover all phases and all description techniques of the SysLab method, and it may not be restricted to a certain class of information processing systems, such as business applications. From that results the requirement that the system model has to be *as general as possible*.

On the other hand, it should be easy to define a semantics based on the system model for the description techniques to be developed. This leads to the requirement that the system model has to be tailored for the description techniques we aim at. This means for instance that we aim at a model supporting the dynamic creation and deletion of components ("objects").

The basic assumption with respect to the structure of information processing systems is that such systems are hierarchically and modularly constructed from a number of components, which may interact in parallel and which can be viewed as information processing systems for themselves. In this case, we call the system a *distributed system*. Distribution here means *spatial distribution* as well as *logical distribution* of functionality across components. However, there are systems which can not be parallelized or distributed any further. Such *basic components* can be modeled using state automata with input and output. The repeated decomposition of a system into subsystems yields a hierarchical system, the structure of which can be viewed as a tree with distributed systems on the inner nodes and with basic components on the leaves.

We are interested in a system model in which each kind of interaction is expressible. In our opinion, each kind of interaction can be viewed as the *exchange of messages* between the interacting components. Such components can be modeled as having *input ports* to receive messages from their environment, and *output ports* to send messages to their environment. Ports constitute the *interface* of a component, they provide the only possibility for the interaction between a component and its environment. The *behavior* of such a component is the relationship between the sequences of messages on its inputs ports and the sequences of messages on its output ports. Systems and their components encapsulate data as well as processes. *Encapsulation of data* means that the state is not directly visible to the environment, but can only be accessed using explicit communication. *Encapsulation of processes* means that the exchange of a message does not imply the exchange of control, and that therefore each component is itself a process.

Exchange of messages between the components of a system is *asynchronous*. This means that a message can be sent independently of a possible readiness of the receiver to receive the message. The requirement for asynchronous communication results from experience in the project FOCUS [BDD$^+$93]: Asynchronous system models provide the most abstract system model for systems with message exchange. They can easily be modeled using stream processing functions, for which a multitude of tractable specification techniques for untimed as well as for timed systems exist ([GS95a], [BDD$^+$93]). Moreover, for stream processing functions a powerful theory for compositional refinement has been developed. By using an asynchronous system model, in contrast to process algebraic approaches like the $\pi$-calculus [Mil91] or CCS [Mil89], we do not have to tackle synchronization issues. To take into account synchronization is in our opinion an issue which is irrelevant in the early phases of system development. However, synchronization can easily be encoded in our model, e.g. by using an appropriate protocol.

If possible, the system model should not impose any constraints concerning the *addressing of messages*. One possibility for addressing is that input and output ports are statically connected through *channels*. Alternatively, it is also possible in our model to address messages using *identifiers*, as they are used in the context of object-oriented programming languages. Moreover, in defining the semantics of object-oriented programming languages we can not assume that the set of components is static, but we have to allow for the dynamic generation of components. These requirements lead to two concepts for communication. The first uses ports and the second uses identifiers. The system model has to be prepared for both communication concepts, where one of them or a combination of both may be chosen if the system model is applied.

However, our system model is not concerned with further object oriented concepts like class descriptions or inheritance hierarchies. These are regarded as description techniques, the semantics of which is defined using the mathematical system model.

To allow for the consideration of systems in which time is relevant, the system model has to provide an explicit notion of time which goes beyond the causality relation formalized e.g. by the monotonicity requirement for stream processing functions [BDD$^+$93]. We assume that a *discrete time*, which is obtained by partitioning the time scale into equidistant time intervals, is sufficient for the purpose of SYSLAB.

# Chapter 3

# Basic Notions

We model an interactive system by a network of autonomous components
which communicate via directed channels in a time-synchronous and message-
asynchronous way. Time-synchrony is achieved by using a global clock split-
ting the time axis into discrete, equidistant time units. Message-asynchrony is
achieved by allowing arbitrary, but finitely many messages to be sent along a
channel in each time unit.

## 3.1 Communication Histories

We model the communication histories of directed channels by infinite streams
of finite streams of messages. Each finite stream represents the communication
history within a time unit. The first finite stream contains the messages received
within the first time unit, the second the messages received within the second
time unit, and so on (see Figure 3.1). Since time never halts, any complete
communication history is infinite.

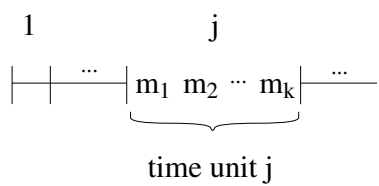

Figure 3.1: Timed stream

Let $D$ be the set of all messages. Then $[D^*]$ is the set of all complete commu-
nication histories, and $(D^*)^*$ is the set of all partial communication histories[1].

---

[1]For an arbitrary set $S$, we denote by $S^*$ the set of finite streams over $S$ and by $[S]$ the set
of infinite streams over $S$.

A *named communication history* is a stream tuple $\iota \in (I \to [D^*])$ assigning to each channel named by the elements of $I$ a complete communication history[2]. Similarly, a *partial* named communication history $\iota \in (I \to (D^*)^*)$ assigns to each channel named by the elements of $I$ a partial communication history. Note that if $I = \emptyset$ then both $I \to [D^*]$ and $I \to (D^*)^*$ contain exactly one named stream tuple, the empty stream tuple and the empty partial stream tuple respectively.

Given a timed stream $s \in [D^*]$ and natural number $j$, $s\downarrow_j$ denotes the prefix of $s$ containing exactly the first $j$ sequences. The $\downarrow$ operator is overloaded to named stream tuples in a point-wise style, i.e. for a named stream tuple $\alpha$, $\alpha\downarrow_j$ denotes the result of applying $\downarrow_j$ to each component of $\alpha$. In an element-wise style, it is also extended to sets of named stream-tuples.

### 3.2 Sum and Projection

Named sequences of messages and named communication histories can be combined and restricted using the overloaded *sum* and *projection* operators. Before defining these operators, let us denote by $\alpha\updownarrow_n \in I \to D^*$ the content of the named communication history $\alpha \in I \to [D^*]$ at time unit $n$.

**Definition 1 (Sum)** *For any $I, J$ such that $I \cap J = \emptyset$ we define the sum on named sequences as follows:*

$$+ \in (I \to D^*) \times (J \to D^*) \to ((I \cup J) \to D^*)$$

$$(\varphi + \psi)(i) \stackrel{def}{=} \begin{cases} \varphi(i) & \text{if } i \in I \\ \psi(i) & \text{if } i \in J \end{cases}$$

*We overload this sum to communication histories such that for all $n$:*

$$+ \in (I \to [D^*]) \times (J \to [D^*]) \to ((I \cup J) \to [D^*])$$

$$(\alpha + \beta)\updownarrow_n \stackrel{def}{=} \alpha\updownarrow_n + \beta\updownarrow_n$$

□

**Definition 2 (Projection)** *The projection of $\theta \in I \to D^*$ on $O$ is written as $\theta|_O$. It is an element of $(I \cap O) \to D^*$ such that:*

$$(\theta|_O)(i) \stackrel{def}{=} \theta(i) \quad \text{for each } i \in (I \cap O)$$

*The projection of $\alpha \in I \to [D^*]$ on a set of names $O$ is also written as $\alpha|_O$. It is an element of $(I \cap O) \to [D^*]$ such that for all $n$:*

$$(\alpha|_O)\updownarrow_n \stackrel{def}{=} (\alpha\updownarrow_n)|_O$$

□

Note that if $I \cap O = \emptyset$ then $\theta|_O$ and $\alpha|_O$ represent the empty named sequence and the empty named stream tuple respectively. Projection is extended to sets of named stream-tuples in an element-wise style.

---

[2] Note that $I \to [D^*]$ and $[I \to D^*]$ are isomorphic and therefore interchangeable.

# Chapter 4

# Basic Components

In this section, we present two views on basic components. In the *state-box* view [HM93], a basic component is modeled using some kind of state machine, so called *timed port automata*. In the *black-box* view, we abstract from the internal structure of a system and model a component solely by its external behavior, which is given by a set of functions mapping input streams to output streams.

## 4.1 State-Box View

There are many automaton models for modeling interactive systems. Examples are I/O-automata [LS89] and automata with output [RK96]. In this paper we use *timed port automata*, which provide an elegant way to model timed and untimed components in our system model, and which can be easily composed.

### 4.1.1 Timed Port-Automata

In the state-based approach, we model components by timed port automata, as defined in [GR95] and similarly in [Rum96]. Their interface consists of a set of input and output ports. To achieve modularity, ports may be hidden.

**Definition 3 (Port signature)** *Let $D$ be a set of* data values *and $I, O, H$ pairwise disjoint sets of* input, output *and* hidden or internal ports *respectively. A port signature is a tuple $\Sigma = (D, I, O, H)$. Given a port signature $\Sigma = (D, I, O, H)$, we denote by $C = I \cup O \cup H$ the set of all ports in $\Sigma$.*  □

As we already pointed out in the introduction, components communicate asynchronously. As a consequence, timed port automata are not allowed to block their environment. Therefore, in every state, they have to react to every possible input. The automata are timed, i.e. each reaction (or transition) takes place in one time unit. The input or output associated with a transition consists of a finite sequence of messages. The empty sequence denotes the absence of any input- or output message.

**Definition 4 (Timed port automaton)** *A* timed port automaton *is a tuple* $A = (\Sigma, S, S^0, \delta)$ *where:*

- $\Sigma$ *is a port signature,*
- *S is a set of* states,
- $S^0 \subseteq S$ *is the set of* start states,
- $\delta \in S \times (C \to D^*) \times S$ *is the transition relation, which is required to be* reactive:

$$\forall s \in S, i \in (I \to D^*) : \exists t \in S, \theta \in (C \to D^*) : (s, \theta, t) \in \delta \land \theta|_I = i.$$

□

Note that reactiveness requires the existence of at least one transition in every state, even if the automaton does not have any input channels. This assures time progress because any transition takes place in exactly one time unit. Another consequence is that a timed port automaton cannot have an empty transition relation.

If $(s, \theta, t) \in \delta$ we also write it as $s \xrightarrow{\theta}_\delta t$ or simply as $s \xrightarrow{\theta} t$ if $\delta$ is clear from the context. A named sequence $\theta \in N \to D^*$ is called an *input action* if $N = I$, an *output action* if $N = O$ and a *hidden action* if $N = H$. Input and output actions are also called *external actions*.

**Example 1 (The timed merge automaton)** *The automaton FM, defined by $FM = (\Sigma, \{s\}, \{s\}, \delta)$, where $\Sigma = (D, \{i, j\}, \{o\}, \emptyset)$, consumes data items from the channels i and j and sends them to o. The automaton is fair, i.e. it never neglects any incoming message indefinitely. The automaton has only one state which is also the initial one. The transition relation is defined as follows[1]:*

$$\delta = \{s \xrightarrow{\{i \mapsto a,\ j \mapsto b,\ o \mapsto c\}} s \ \mid \exists p \in \{i, j\}^{\#c} : \ a = pr_i(p, c) \land b = pr_j(p, c)\}$$

where

$$pr_k(k\ \&\ p, m\ \&\ a) = m\ \&\ pr_k(p, a)$$
$$pr_k(l\ \&\ p, m\ \&\ a) = pr_k(p, a)$$
$$pr_k(\epsilon, \epsilon) = \epsilon, \qquad \text{for } k, l \in \{i, j\} \text{ and } k \neq l$$

*Each transition corresponds to a merge of a and b. Note that if a or b is $\epsilon$ then c is b or a, respectively.* □

### 4.1.2 Executions, Schedules and Behaviors

**Definition 5 (Execution, schedule, behavior)** *An* execution *of an automaton A is an infinite sequence $s^0, \theta^0, s^1, \theta^1, \ldots$ such that $s^0 \in S^0$ and $\forall i : s^i \xrightarrow{\theta^i} s^{i+1}$. We denote the set of all executions of A by $execs(A)$.*

---

[1] $\#c$ is the length of sequence $c$; $m\ \&\ a$ appends the message $m$ in front of sequence $a$.

*The* schedule *sched($\alpha$) of an execution $\alpha$ is the subsequence of $\alpha$ containing only* actions *in $\alpha$. We denote the set of schedules of $A$ by sched($A$).*

*The* behavior *beh($\alpha$) of an execution or schedule $\alpha$ is the subsequence of $\alpha$ containing only* external actions. *We denote by behs($\alpha$) the set of all behaviors of $\alpha$.* □

Note that schedules and behaviors are named communication histories. Given an automaton $A$ and an input stream-tuple $\iota \in I \to [D^*]$, we denote the set of behaviors of $A$ with input $\iota$ by $A[\iota]$ and write it simply $[\iota]$ when $A$ is clear from the context. Formally:

$A[\iota] = \{\alpha \in behs(A) \mid \alpha|_I = \iota\}$

For a deterministic automaton $A[\iota]$ is a singleton for each $\iota$.

### 4.1.3 Strongly Pulse Driven Automata

Timed port automata have a characteristic property: *they process their input incrementally.* In other words, at any moment of time, their output does not depend on future input. This property is called *pulse-drivenness* and it has two variations: *strong pulse-drivenness* and *weak pulse-drivenness*.

The output produced in time unit $t$ by a strongly pulse-driven automaton is not only independent of future input but also of input received in the same time unit. The output produced by a weakly pulse-driven automaton in time unit $t$ can also depend on the input received in time unit $t$. Hence, the strongly pulse-driven automata introduce a delay between input and output while the weakly pulse-driven automata may not.

**Definition 6 (Pulse driven automata)** *An timed port automaton $A$ is called* strongly pulse-driven *iff*

$\forall \iota, \kappa, n : \iota\downarrow_n = \kappa\downarrow_n \Rightarrow [\iota]|_O\downarrow_{n+1} = [\kappa]|_O\downarrow_{n+1}$

*An automaton $A$ is called* weakly pulse-driven *iff*

$\forall \iota, \kappa, n : \iota\downarrow_n = \kappa\downarrow_n \Rightarrow [\iota]|_O\downarrow_n = [\kappa]|_O\downarrow_n$ □

For a deterministic automaton, weak pulse-drivenness says that we can unfold the automaton into a tree of infinite broadness and depth, whose nodes are marked by states and whose branches are marked with input, output and hidden messages $(C \to D^*)$, that lead from the father's state to the son's state.

Due to reactivity of an automaton, each node in the unfolded tree has at least one branch for every input of $I \to D^*$, due to determinism it has at most one branch for every input of $I \to D^*$. Each path in the tree corresponds to a behavior $\alpha$ and each level $n$ corresponds to the set of all behaviors $\alpha\downarrow_n$.

For a nondeterministic automaton, the same unfolding is possible, but now each node in the unfolded tree may have more than one branch for a given input.

**Theorem 1** *Every timed port automaton is weakly pulse-driven.*

**Proof:**   An immediate consequence of reactiveness.   □

**Theorem 2** *The timed merge automaton TMA is weakly pulse-driven but not strongly pulse-driven.*

**Proof:**   Trivial   □

Sometimes we want to require strong pulse-drivenness only on particular subsets of the input and output channels.

**Definition 7 (Strong pulse-drivenness with respect to $(G, P)$)** *A timed port automaton A is strongly pulse-driven with respect to $(G, P)$, where $G \subseteq I$ and $P \subseteq O$, iff*

$$\forall \iota, \kappa, n : \ (\iota|_G)\downarrow_n = (\kappa|_G)\downarrow_n \ \land \ (\iota|_{I\setminus G})\downarrow_{n+1} = (\kappa|_{I\setminus G})\downarrow_{n+1}$$
$$\Rightarrow \ ([\iota]\,|_P)\downarrow_{n+1} = ([\kappa]\,|_P)\downarrow_{n+1}$$

*Obviously, if $P = \emptyset$ then each automaton is strongly pulse-driven wrt. $(G, P)$.*
□

An example of a strongly pulse-driven automaton is the buffer given below.

**Example 2 (A buffer with restricted delay)** *The buffer automaton*

$$BUF = (\Sigma, D^*, \{\epsilon\}, \delta) \quad \text{where} \quad \Sigma = (D, \{i\}, \{o\}, \emptyset)$$

*consumes data items from i and reproduces them with a finite delay on o. The order of the incoming messages is preserved and the buffer capacity is unrestricted. The contents of the buffer is modeled by the set of states $D^*$. To ensure that every received message is also sent, we enforce BUF to send at least one message if the contents of the buffer is not empty. The transition relation is defined below, where $\&$ denotes the concatenation operation on finite sequences:*

$$\delta = \{a \,\&\, s \stackrel{\{i \mapsto b,\ o \mapsto a\}}{\longrightarrow} s \,\&\, b \mid (a, b \in D^*) \land (a \,\&\, s \neq \epsilon \Rightarrow a \neq \epsilon)\}$$

*Note that the delay of every message is bounded by the input received earlier.* □

## 4.2  Black-Box View

In many cases, for methodical reasons one is interested in a more abstract view of a component than the view given by a state machine or by a network (see the following section). In such cases, the black-box view, also called the history-based approach, can be used [GS95b, GR95]. In the history-based approach we model components by sets of functions.

Each function has the form $f \in (I \to [D^*]) \to (O \to [D^*])$. It maps named input histories to named output histories. The names of the input channels build the set $I$ and the names of the output channels build the set $O$.

The reason for working with infinite histories is that if no action is communicated along an input channel within a time unit, then an empty message sequence occurs in the input history. The lack of this timing information causes the fair merge anomaly [Kel78].

The functions should behave similar to deterministic automata, i.e., they should process their input incrementally.

**Definition 8 (Pulse driven functions)** *Stream processing functions whose output until time j (j + 1) is completely determined by the input until time j are called* weakly (strongly) pulse-driven. *Formally:*

$\forall \iota, \kappa, j : \iota\downarrow_j = \kappa\downarrow_j \Rightarrow f(\iota)\downarrow_j = f(\kappa)\downarrow_j$     *(weakly pulse-driven)*
$\forall \iota, \kappa, j : \iota\downarrow_j = \kappa\downarrow_j \Rightarrow f(\iota)\downarrow_{j+1} = f(\kappa)\downarrow_{j+1}$     *(strongly pulse-driven)*

*We use the arrow $\rightarrow$ for sets of strongly pulse-driven functions and the arrow $\stackrel{w}{\rightarrow}$ for sets of weakly pulse-driven functions.* □

Strongly and weakly pulse-driven functions correspond to contractive and non-expansive functions in the metric of streams (see [GS95b, GR95]). As a consequence, by Banach's fixed point theorem, strong pulse-drivenness guarantees unique fixed points of feedback loops.

**Theorem 3** *Projection is a pulse-driven function. Sum is a pulse-driven function in each argument.*

**Proof:**    $(\iota + \kappa)\downarrow_n$ and $(\iota|_O)\downarrow_n$ only depend on $\iota\downarrow_n$ and $\kappa\downarrow_n$. □

To allow non-deterministic components, we model components by sets of weakly pulse-driven functions:

**Definition 9 (Components)** *A component whose input and output channels are named by I and O, respectively, is given by a nonempty set of weakly pulse-driven functions*

$F \subseteq (I \rightarrow [D^*]) \stackrel{w}{\rightarrow} (O \rightarrow [D^*]),$

*that is* closed *in the sense that for all weakly pulse-driven functions f of the same signature*

$(\forall \iota \in (I \rightarrow [D^*]) : \exists f' \in F : f(\iota) = f'(\iota)) \Rightarrow f \in F.$ □

The above definition is very powerful. It not only makes the model *fully abstract*, but it also allows us to handle *unbounded nondeterminism*. Note that the use of relations instead of sets of functions is problematic in connection with unbounded nondeterminism (see for example [Cos85, NP92]).

# Chapter 5

# Networks

A component may be distributed in a network of subcomponents. In this case a third view, the so called network-view describes the distribution of a component. It is given by the set of subcomponents as well as their connections. Formally, connections are described by named streams. Two components sharing a channel as input resp. output stream are connected together. Moreover, in this section we show how one can derive the state-box as well as the black-box view of a distributed component if the corresponding views of its subcomponents are given.

## 5.1 Composition of State-Box Views

In this section, we first show how two automata can be composed into a new automaton. This result is then generalized to the composition of infinite sets of automata. Moreover, we introduce a hiding operator which provides control over the scoping of channels.

### 5.1.1 Binary Composition

When composing automata the first to decide is, whether automata should be allowed to write on the same channels or not. In the case of the one-to-many composition only one automaton is allowed to write on a given channel. This assures that no merging of messages is necessary. We also require that hidden channels are private.

**Definition 10 (Compatible port signatures)** *The signatures $\Sigma_1$ and $\Sigma_2$ are called* compatible *iff $O_1 \cap O_2 = H_1 \cap C_2 = H_2 \cap C_1 = \emptyset$* □

If two automata are composed, the output channels of one automaton are connected to the input channels with the same name of the other automaton. A graphical illustration is given in Figure 5.1. The set of hidden channels of the composed automaton is the union of the sets of hidden channels of the components.

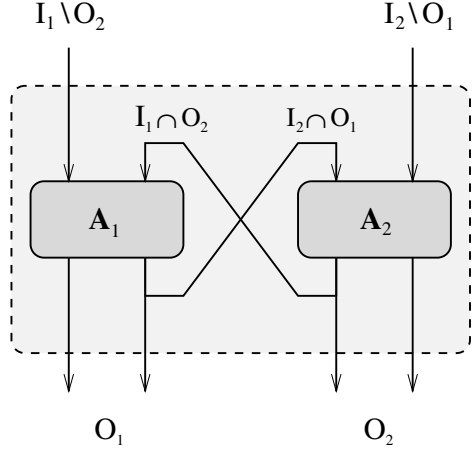

Figure 5.1: Binary composition

**Definition 11 (Binary composition of signatures)** *If $\Sigma_1$ and $\Sigma_2$ are compatible then the sets*

$$I = (I_1 \setminus O_2) \cup (I_2 \setminus O_1), \quad O = O_1 \cup O_2, \quad H = H_1 \cup H_2$$

*are pairwise disjoint. We may therefore define the* composition *of $\Sigma_1$ and $\Sigma_2$ to be the signature $\Sigma_1 \otimes \Sigma_2 = (D, I, O, H)$. As before $C = I \cup O \cup H$.* □

The main difficulty of untimed approaches is composition. The absence of information about the causality or the relative timing of the messages exchanged by two automata leads to the Brock-Ackerman anomaly. The I/O automata approach introduces causality between messages by arranging them in a linear order: the synchronized merge of the messages of each automaton. Synchronization takes place when one automaton sends a message which the other automaton has to accept due to its input enabledness ([LS89], [BDDW91]). To make interleaving possible, each transition of an I/O automaton is marked with *only one* message: either an input or an output message. As a consequence, this approach is not explicitly concurrent. This is the case even if one would introduce special time transitions.

We work with a *global clock*: each transition takes place in the same, constant unit of time in both automata. Hence, our approach is *time-synchronous* and *message-asynchronous*. By using a global clock, we have enough causality information to obtain a very simple yet powerful notion of composition: the set of input and output messages of the composed automaton at a given time $n$ is simply the union of the input and output messages of the components at the same time $n$. Note that union also takes care for communication. If an output channel of one automaton is connected to an input channel of the other automaton, the input channel does not appear in the composed automaton.

**Definition 12 (Binary composition of automata)** *Given two timed port automata $A_1$ and $A_2$ with compatible signatures. Then the* binary composition *of $A_1$ and $A_2$ is $A_1 \otimes A_2 = (\Sigma, S, S^0, \delta)$ which is defined as follows:*

- $\Sigma = \Sigma_1 \otimes \Sigma_2$,
- $S = S_1 \times S_2$,
- $S^0 = S_1^0 \times S_2^0$,
- $\delta_1 \otimes \delta_2 \in S \times (C \to D^*) \times S$
  $\delta_1 \otimes \delta_2 = \{(s_1, s_2) \xrightarrow{\theta} (t_1, t_2) \mid \quad s_1 \xrightarrow{\theta|_{C_1}} t_1 \in \delta_1 \ \wedge \ s_2 \xrightarrow{\theta|_{C_2}} t_2 \in \delta_2\}$ □

Similarly to the composition for I/O automata, this composition allows automata to block each other, e.g., the composition can have an empty transition relation.

**Example 3 (Blocking)** *Given two automata with complementary signatures*

$A_1 = (\Sigma_1, \{s_1\}, \{s_1\}, \delta_1), \quad \Sigma_1 = (\mathbb{N}, \{i\}, \{o\}, \emptyset)$
$A_2 = (\Sigma_2, \{s_2\}, \{s_2\}, \delta_2), \quad \Sigma_2 = (\mathbb{N}, \{o\}, \{i\}, \emptyset)$

*and with the following transition relations, where $1 \& a$ appends the natural number 1 in front of the sequence $a$:*

$\delta_1 = \{(s_1, \{i \mapsto a, \ o \mapsto 1 \& a\}, s_1) \mid a \in \mathbb{N}^*\}$
$\delta_2 = \{(s_2, \{o \mapsto a, \ i \mapsto 1 \& a\}, s_2) \mid a \in \mathbb{N}^*\}$

*The composed automaton*

$A = (\Sigma_1 \otimes \Sigma_2, \{(s_1, s_2)\}, (s_1, s_2), \delta_1 \otimes \delta_2), \quad \Sigma_1 \otimes \Sigma_2 = (\mathbb{N}, \emptyset, \{i, o\}, \emptyset)$

*has an empty transition relation $\delta_1 \otimes \delta_2$, because there is no $\theta = \{i \mapsto a, \ o \mapsto b\}$ such that $b = 1 \& a$ as required by the first automaton and that $a = 1 \& b$ as required by the second automaton. This can be interpreted either as divergence or as blocking. In the second case each automaton waits without success for an output produced by the other automaton.* □

Since $\delta_1 \otimes \delta_2$ is empty, $A$ is not a timed port automaton. This means that composition is a partial operation. However, if the output produced by one automaton is independent of the output produced by the other automaton then the automata cannot block each other and the composition is well defined.

**Theorem 4** *Given two timed port automata $A_1$ and $A_2$. Let $G = I_1 \cap O_2$ and $P = I_2 \cap O_1$. If either $A_1$ is strongly pulse-driven with respect to $(G, P)$ or $A_2$ is strongly pulse-driven with respect to $(P, G)$, then $\delta_1 \otimes \delta_2$ is well defined, that means $A_1 \otimes A_2$ is a port automaton.*

The proof for this and the following two theorems can be seen in [GR95]. In the sequel we always assume that the conditions required by the above theorem hold. Note that if any of $G$ or $P$ is empty, the composition is well defined even if both automata are only weakly pulse-driven.

**Theorem 5** *Given timed port automata $A_1$ and $A_2$. Then*

$$\begin{aligned}
execs(A_1 \otimes A_2) &= \{e \mid e|_{A_1} \in execs(A_1) \wedge e|_{A_2} \in execs(A_2)\} \\
scheds(A_1 \otimes A_2) &= \{e \mid e|_{A_1} \in scheds(A_1) \wedge e|_{A_2} \in scheds(A_2)\} \\
behs(A_1 \otimes A_2) &= \{e \mid e|_{A_1} \in behs(A_1) \wedge e|_{A_2} \in behs(A_2)\}
\end{aligned}$$

Strong pulse-drivenness is preserved by composition.

**Theorem 6** *If $A_1$ and $A_2$ are strongly pulse-driven automata then so is $A_1 \otimes A_2$.*

### 5.1.2 Infinite Composition

We now extend these results to the composition of infinitely many components. Infinitely many components are used to model dynamic creation and destruction of components in the system model. For example, the infinite set of all instances of a class can be represented by an infinite set of components in the system model. These objects are created and deleted by sending appropriate messages to them.

Observe that infinite composition cannot be done by a generalisation of binary composition to arbitrary (finite) composition, because there is a limit process involved.

**Definition 13 (Compatible port signatures)** *Given a countable set $J$. The signatures $(\Sigma_j)_{j \in J}$ are called* compatible *iff*

$$O_j \cap O_k = H_j \cap C_k = H_k \cap C_j = \emptyset$$

*for each $j, k \in J, j \neq k$.*  □

Observe that while output channels are private, input channels may be shared among signatures.

**Definition 14 (Infinite composition of signatures)** *Given a compatible set of signatures $(\Sigma_j)_{j \in J}$. Then the sets*

$$I = (\bigcup_{j \in J} I_j) \setminus (\bigcup_{j \in J} O_j), \quad O = \bigcup_{j \in J} O_j, \quad H = \bigcup_{j \in J} H_j$$

*are pairwise disjoint. We may therefore define the* composition *of $(\Sigma_j)_{j \in J}$ to be the signature $\otimes_{j \in J} \Sigma_j = (D, I, O, H)$. As before $C = I \cup O \cup H$.*  □

**Definition 15 (Infinite composition of automata)** *Given the set of timed port automata $(A_j)_{j \in J}$ with compatible signatures. The* infinite composition of *$(A_j)_{j \in J}$ is*

$$\otimes_{j \in J} A_j = (\Sigma, S, S^0, \delta)$$

*which is defined as follows:*

- $\Sigma = \otimes_{j \in J} \Sigma_j$,
- $S = \times_{j \in J} S_j$,
- $S^0 = \otimes_{j \in J} S_j^0$,
- $\otimes_{j \in J} \delta_j \in S \times (C \to D^*) \times S$
  $$\otimes_{j \in J} \delta_j = \{s \xrightarrow{\theta} t \mid \forall j \in J: \; s(j) \xrightarrow{\theta|_{C_j}} t(j) \in \delta_j\}$$

□

**Theorem 7** *Given the timed port automata $(A_j)_{j \in J}$ with compatible signatures. Let $I_j^f = I_j \cap O$ and $O_j^f = O_j \cap (\bigcup_{k \in J} I_k)$. If every component $A_j$ is strongly pulse-driven with respect to $(I_j^f, O_j^f)$, then $\otimes_{j \in J} \delta_j$ is well defined, i.e., $\otimes_{j \in J} A_j$ is a timed port automaton.*

**Proof:** We have to show that $\otimes_{j \in J} \delta_j$ is reactive. Given $s = (s_j)_{j \in J}$ and $i \in I \to D^*$. Since each $A_j$ is strongly pulse-driven wrt. $(I_j^f, O_j^f)$, the output on $O_j^f$ at a time unit does not depend on its input on the feedback channels $I_j^f$ at this time unit.

This means, the output $o^f \in O \to D^*$ such that $o^f|_{A_j} = o_j^f$ for all $j \in J$ is completely determined by the state $s$ and the input $i$. Since $I_j^f \subseteq \bigcup_{k \in J} O_k^f$, the input for $A_j$ is fixed and given by $(i + o^f)|_{I_j}$. By the reactivity of $A_j$, we can find a state $t_j$ and a $\theta_j$ such that $\theta_j|_{I_j} = (i + o^f)|_{I_j}$ and $(s(j), \theta_j, t_j) \in \delta_j$. Take $t = (t_j)_{j \in J}$ for the next state and $\theta$ such that $\theta|_{A_j} = \theta_j$ for the action. □

The obvious generalization of theorem 5 to infinite composition is left to the reader.

### 5.1.3 Hiding

Hiding is a very important operation which provides control over the scoping of channels. This is of great importance in the modular development of reactive systems.

There are two reasons for using a separate hiding operator instead of a composition which automatically hides the interconnected channels. First, the composition with hiding is not associative. Second, our compositionality result for behaviors could not have been formulated in the simple way we did, because the component behaviors would have contained information which were not present in the behavior of the composed automata.

**Definition 16 (Hiding)** *Given a timed port automaton*

$$A = (\Sigma, S, S^0, \delta), \quad \text{where} \quad \Sigma = (D, I, O, H)$$

*and a set $P \subseteq O$. The automaton $\nu P : A$ is then defined as follows:*

$$\nu P : A = (\Sigma', S, S^0, \delta), \quad \text{where} \quad \Sigma' = (D, I, O \setminus P, H \cup P)$$

□

It is easy to see that $\nu P : A$ is a timed port automaton.

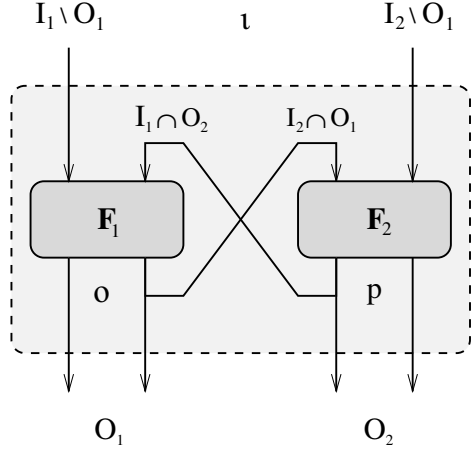

Figure 5.2: Binary composition

## 5.2 Composition of Black-Box Views

In the preceding section we studied how components modeled by automata can be composed. Because in many cases components are not modeled using automata, but using the black-box view, it is also important that one can compose such black-box components.

### 5.2.1 Binary Composition

We define the composition slightly more general than in [GS95b] because we relax the pulse-drivenness requirements and allow one-to-many communication.

**Definition 17 (Binary composition)** *Given two components $F_1$ and $F_2$ with input channels in $I_1$ and $I_2$ and output channels in $O_1$ and $O_2$:*

$$F_1 \subseteq (I_1 \to [D^*]) \xrightarrow{w} (O_1 \to [D^*]), \qquad F_2 \subseteq (I_2 \to [D^*]) \xrightarrow{w} (O_2 \to [D^*])$$

*such that $I_1 \cap O_1 = I_2 \cap O_2 = O_1 \cap O_2 = \emptyset$. Let $J = I_1 \cap O_2$ and $P = I_2 \cap O_1$.*

*If either $F_1$ is strongly pulse-driven with respect to $(J, P)$ or $F_2$ is strongly pulse-driven with respect to $(P, J)$ then we may compose them in accordance with Figure 5.2. We refer to the resulting network as the* binary *composition of $F_1$ and $F_2$ and write it as $F_1 \otimes F_2$. Formally:*

$$I = (I_1 \setminus O_2) \cup (I_2 \setminus O_1), \qquad O = O_1 \cup O_2,$$

$$F_1 \otimes F_2 \subseteq (I \to [D^*]) \xrightarrow{w} (O \to [D^*]),$$

$$F_1 \otimes F_2 = \{f \mid \forall \iota : \exists f_1 \in F_1, f_2 \in F_2 :$$
$$f(\iota) = o + p \text{ where } o = f_1((\iota + p)|_{I_1}), \ p = f_2((\iota + o)|_{I_2})\}$$
□

**Theorem 8** $F_1 \otimes F_2$ *is a component.*

**Proof sketch:** That $F_1 \otimes F_2 \neq \emptyset$ follows straightforwardly from Banach's fixed point theorem. Closedness follows from the nesting of the quantifiers ($\exists f_1, f2 \ldots$ inside of $\forall \iota$). □

### 5.2.2 Infinite Composition

**Definition 18 (Infinite composition)** *Given the components* $(F_j)_{j \in J}$ *with input channels in* $I_j$ *and output channels in* $O_j$,

$$F_j \subseteq (I_j \to [D^*]) \xrightarrow{w} (O_j \to [D^*]),$$

*such that* $I_j \cap O_j = O_j \cap O_k = \emptyset$ *for each* $j, k \in J, j \neq k$. *Let* $I_j^f = I_j \cap (\bigcup_{j \in J} O_j)$ *and* $O_j^f = O_j \cap (\bigcup_{j \in J} I_j)$ *be the feedback channels of* $F_j$.

*If each* $F_j$ *is strongly pulse-driven with respect to* $(I_j^f, O_j^f)$, *then the infinite composition* $\otimes_{j \in J} F_j$ *is defined as follows:*

$$I = (\bigcup_{j \in J} I_j) \setminus O, \qquad O = \bigcup_{j \in J} O_j,$$

$$\otimes_{j \in J} F_j \subseteq (I \to [D^*]) \xrightarrow{w} (O \to [D^*]),$$

$$\otimes_{j \in J} F_j = \{f \mid \forall \iota : \forall j \in J : \exists f_j \in F_j :$$
$$f(\iota)|_{O_j} = f_j((\iota + o)|_{I_j}) \text{ where } o = f(\iota)$$

□

**Theorem 9** *Given components* $(F_j)_{j \in J}$, *then* $\otimes_{j \in J} F_j \neq \emptyset$.

**Proof:** Since $F_j$ is a component for each $j$, we may find functions $f_j$ such that $f_j \in F_j$. We define function $g$ such that:

$$g \in (O \to [D^*]) \times (I \to [D^*]) \to (O \to [D^*])$$
$$g(p, \iota)|_{O_j} = f_j((\iota + p)|_{I_j})$$

where $p \in (O \to [D^*])$. As composition of all $f_j$ is $g$ contractive ([GR95]) in each $O_j$. Since this definition is equal to $g(p, \iota) = \Sigma_{j \in J} g(p, \iota)|_{O_j}$, $g$ is contractive in $p$. By Banach's fixed-point theorem, $f = \mu g$ is well defined, yielding a function:

$$f \in (I \to [D^*]) \to (O \to [D^*])$$

which is by construction $f \in \otimes_{j \in J} F_j$. □

**Theorem 10** $\otimes_{j \in J} F_j$ *is a component.*

**Proof:** By theorem 9 it is enough to prove that $\otimes_{j \in J} F_j$ is closed. Suppose $f \in (I \to [D^*]) \xrightarrow{w} (O \to [D^*])$ and

$$\forall \iota : \exists f' \in \otimes_{j \in J} F_j : f(\iota) = f'(\iota).$$

Then for a given $\iota$ and $f'$ there are functions $f_j \in F_j$ such that:

$$f'(\iota)|_{O_j} = f_j((\iota + o)|_{I_j}) \text{ where } o = f'(\iota).$$

Hence, for every $\iota$ there are functions $f_j \in F_j$ such that:

$$f(\iota)|_{O_j} = f_j((\iota + o)|_{I_j}) \quad \text{where} \quad o = f(\iota).$$

In other words $f \in \otimes_{j \in J} F_j$. □

### 5.2.3 Hiding

As for timed port automata, hiding provides control over the scoping of channels. This is of great importance in the modular development of reactive systems.

**Definition 19 (Hiding)** *Given a component $F$ with input channels in $I$ and output channels in $O$*

$$F \subseteq (I \to [D^*]) \xrightarrow{w} (O \to [D^*])$$

*and a set of channel names $P \subseteq O$. The component $\nu P : F$ is defined as follows:*

$$\nu P : F \subseteq (I \to [D^*]) \xrightarrow{w} ((O \setminus P) \to [D^*])$$
$$\nu P : F = \{f \mid \forall \iota : \exists g \in F : f(\iota) = g(\iota)|_{O \setminus P}\}$$

□

**Theorem 11** *Given a component $F$ with inputs in $I$ and outputs in $O$ and a set $P \subseteq O$. Then $\nu P : F$ is also a component.*

**Proof:** Trivial. □

# Chapter 6

# System Model

In the previous sections we presented techniques for modeling different views of a system and for relating these views. We will now show how the SysLab system model can be defined using these techniques. In comparison with [RKB95], we have also slightly extended the model, because we now also allow the duplication of messages and broadcast communication.

## 6.1 Outline of the System Model

The SysLab system model serves as an underlying semantic model mainly used to define the semantics of description techniques used in the SysLab method. It is itself based on pure mathematics. The system model aims at describing general *information processing systems*, not just focusing for instance on hardware- or database systems.

The system model is *hierarchical* and *modular*. It is hierarchically composed of *interacting components*, where the leafs in the hierarchy are called *basic components* and internal nodes are called *distributed components*. All components interact just by *message passing* via *input* and *output ports*. Data (states) of components are encapsulated, no sharing occurs. Messages between components are sent *asynchronously*. No addressing concept for messages is coded in the system model. This allows for using identifiers as e.g. in object-oriented systems as well as using direct connections as e.g. in hardware systems. Dynamic creation of components can be coded in our model by using possibly infinite sets of components, which can be "created" and "deleted" by special messages.

We do not abstract from time, because this on the one hand allows us to describe real-time systems and on the other hand prevents semantic problems as e.g. the merge anomaly. We use a *discrete linear time* by dividing infinite communication histories into infinitely many intervals of the same length.

## 6.2 Component Hierarchy

If we adopt a glass-box view when looking at a real system we find a hierarchy of components which are conceptually or spatially distributed.

To describe this view we attach a unique *identifier* to every component in the system. We therefore use the enumerable set of identifiers $ID$. To describe the system hierarchy we use the function

$$Parts : ID \to \mathcal{P}(ID)$$

where $Parts(i)$ denotes the identifiers of all sub-components of component $i$.

A component $c$ is a basic component exactly if $Parts_c = \emptyset$. We distinguish between the set of basic components $ID_b$ and the set of distributed components $ID_d$. Both sets are a partition of the total set of identifiers, i.e. they are disjoint and their union is the total set of identifiers $ID$.

Note that *Parts* must denote a tree, but this tree may be infinitely broad and deep to model dynamic processes. A special identifier $RootSystem \in ID$ denotes the root of this tree.

## 6.3 Signature of Components

Every component with identifier $c \in ID$ has a set of input and output ports which are used to receive and send messages to other components. These ports build the *signature* of a component. Given an enumerable set of *portnames* $P$ we attach input ports $In_c$ and output ports $Out_c$ to every component $c \in ID$ by

$$\begin{aligned} In &: ID \to \mathcal{P}(P) \\ Out &: ID \to \mathcal{P}(P) \end{aligned}$$

Every port is connected to at most one component:

$$\begin{aligned} c \neq d &\Rightarrow (In_c \cup Out_c) \cap (In_d \cup Out_d) = \emptyset \\ In_c \cap Out_c &= \emptyset \end{aligned}$$

## 6.4 Behavior of Components

Besides its signature, every component has a behavior. We allow our components to be nondeterministic. This is modeled by using a set of stream processing functions with appropriate signature for every component:

$$Behavior_{c \in ID} \subseteq ((In_c \to [D^*]) \xrightarrow{w} (Out_c \to [D^*]).$$

As basic components define their behavior on their own, the behavior of distributed components is composed of the behaviors of their parts and of the behavior of the communication medium. The communication medium will be introduced later for modeling the transmission of messages among components and their ports.

## 6.5 Basic Components

A basic component is not distributed. Its behavior can be given either in a property oriented way as a black-box or in a state oriented way as a state-box. Given a global set $ST$ of states, we assign a subset of these states to every basic component:

$$States : ID_b \to \mathcal{P}(ST).$$

The set of states is usually determined by data definitions. For example in object-oriented frameworks, these states are made of attribute definitions. Moreover, under some circumstances we may also encode a control part in the state set.

A set $State_c^0$ contains the initial states of the component $c$:

$$State_c^0 \subseteq States_c$$

To describe the behavior of a basic component $c$, a state transition relation $\delta_c$ is used. It is allowed to describe nondeterminism as well as underspecification of the basic component. So we have a nondeterministic transition relation of signature:

$$\delta_c : States_c \times (In_c \cup Out_c \to D^*) \times States_c$$

defining at least one pair of successor state and output for every state and every input on the input ports. Together with the signature, the initial state and the state set the transition relation forms a timed port automaton $A_c$

$$A_c = ((In_c, Out_c, \emptyset), States_c, State_c^0, \delta_c).$$

As described in [GR95], every port automaton $A_c$ generates a set of weakly pulse-driven stream processing functions $[A_c]$. This set determines the behavior of component $c$:

$$c \in ID_b \Rightarrow Behavior_c = [A_c].$$

## 6.6 State of Basic Components

The above mentioned global set of states $ST$ comprises the data as well as the control state of each component. The data state is usually described by some kind of algebra, resp. its sorts. In an implementation the control state is given by some kind of program counter. In the more abstract system model, we may represent a control state by a prophecy variable determining future output together with the data state, that will be reached, when all the prophecy output is done. Thus, internally we do not stutter the reaction of a component to some input stimulus by subsequent internal transition on data states, but process the final data state immediately and keep the resulting output messages in an output buffer as prophecy. This allows us to deal only with consistent sets of data states and to have control states represented.

As the input of a component is timed, we have to accept input messages at every time. To get a further degree of description freedom we decide to add an input buffer for not yet processed input messages and thus get a state space similar to [GKRB96]. The state space of every basic component $c \in ID_b$ can be seen as:

$$State_c = (In_c \to D^*) \times Data_c \times (Out_c \to D^*)$$

and the initial state space as:

$$State_c^0 \subseteq \{\epsilon\} \times Data_c \times \{\epsilon\}$$

where $Data_c$ is the data space of a basic component.

## 6.7 Sorts in the System Model

The data states of basic components and the input and output channels of every component are usually typed. Therefore, we associate a sort of an algebra with every

- channel $p \in P$ (and therefore its messages) and every
- space of data states $Data_c$ of basic components $c \in ID_b$.

Thus, we assume the existence of an underlying algebra that provides sorts for all necessary sets of the system model. Algebras can be described conveniently using abstract data types.

## 6.8 Communication Medium

A distributed component $c$ is made of at least one part (subcomponent). Every kind of computation of $c$ is performed within these parts. However, the connection structure of these parts and the strategy of message deliverance is not coded in its parts, but in an additional "component" called *communication* *medium*. This communication medium is also responsible to deliver messages from and to the environment of a component. It may therefore be considered as a membrane between the parts of a component as well as between the interior and the environment of a component. The communication medium has to obey several restrictions:

- Messages with same source and destination port have to maintain order,
- no messages are generated,
- messages are not modified,
- the destinations of a message do not depend on the state of the communication medium,
- every destination of a message receives a message exactly once.

In contrast to the system model in [RKB95], we now require the communication medium to duplicate messages, if there is more than one destination for a message. This generalization allows us to model *broadcasting*, as well as the wiring of hardware, where one output channel is often connected to several input channels.

In principle the communication medium is allowed to delay messages, but not infinitely long. If one is interested in specifying real-time systems, the medium may be refined to a medium without delay. Although the medium is not necessarily a component that is intended to be implemented, we may describe the communication medium as an ordinary component. It has a signature and a behavior and because of the possible delay also an internal state. In Figure 6.1 a network view shows how a component $c$ is decomposed into the communication medium $CM_c$ and the parts of $c$.

The communication medium of the distributed component $c$ is modeled as component $CM_c$. As signature, it has all the output ports of the parts of $c$ as input ports and vice versa and the input- and output ports of component $c$ itself. We define:

$OutParts_c = \bigcup_{d \in Parts_c} Out_d$
$InParts_c = \bigcup_{d \in Parts_c} In_d$

The communication medium $CM_c$ then has the following signature, defined by using two further definitions $Origins_c$ for the input channels and $Destination_c$ for the output channels of $CM_c$.

$Origins_c \quad = In_c \ \cup OutParts_c$
$Destinations_c = Out_c \cup InParts_c$
$CM_c \subseteq (Origins_c \to [D^*]) \xrightarrow{w} (Destinations_c \to [D^*])$

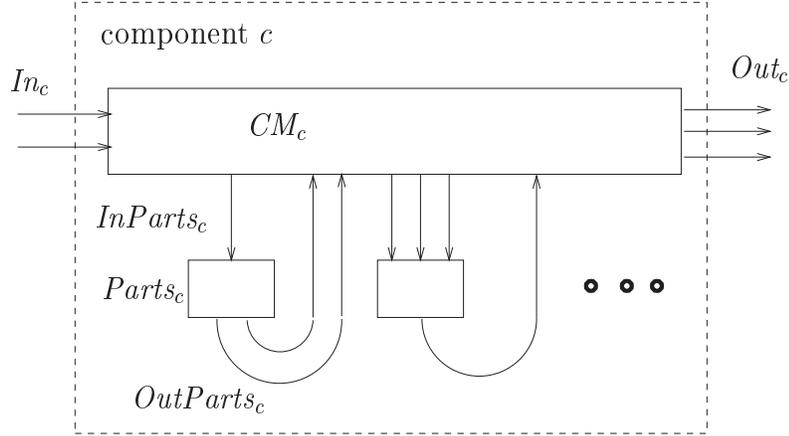

Figure 6.1: Distribution of a component

The behavior of the communication medium can now be described by a timed port automaton. This description gives another view on $CM_c$ but allows basically the same behavior as described in [RKB95], generalized by the new duplication mechanism for messages. The communication medium is modeled as a set of buffers, containing the messages that are still to be delivered. The first purpose of the communication medium is therefore to distribute the incoming messages into the proper buffers. As the destination of a message does not depend on the internal state of the communication medium, a function $destination_c$ will be defined to determine the set of destinations of a message within communication medium $CM_c$. Similarity, $origin_c$ determines the unique origin of a message:

$$destination_c: D \to \mathcal{P}(Destinations_c)$$
$$origin_c \quad\ : D \to Origins_c$$

Note that the functions $destination_c$ and $origin_c$ depend on the component $c$ in which they are used for message distribution. This is all we require about these two functions in the system model. If the system model is used, $destination_c$ and $origin_c$ have to be specialized further by stating additional properties. For instance, to prevent messages from being lost, we may require $destination_c$ to be not empty for any message:

$$\forall c \in ID_d, d \in D.\ destination_c(d) \neq \emptyset$$

If exactly one destination for every message is required, we may in addition state:

$$\forall c \in ID_d, d \in D. \, |destination_c(d)| = 1$$

Function $distribute_c$ collects the incoming messages at one time interval and distributes them to the proper buffers:

$$distribute_c : (Origins_c \to D^*) \to (Destinations_c \to D^*)$$
$$\forall d \in Destinations_c, o \in Origins_c, \theta.$$
$$Filter(\{m|origin_c(m) = o\}, distribute_c(\theta).d) =$$
$$Filter(\{m|d \in destination_c(m)\}, \theta.o)$$

The application of $Filter(M, s)$ removes all messages not in $M$ from the streams $s$. This way, the above formula expresses the requirement that all messages from all origins are put in the appropriate buffers for their destination.

Now the timed port automaton for the communication medium can be defined. We give a very general definition that can be specialized to various kinds of communication media. Note that if the set $Origins_c$ is not finite, $distribute_c$ is a partial function. In this case it may happen that within one time unit infinitely many components send a message to the same receiver. To prevent this problem, it is necessary to ensure that at every moment of time just a finite subset of components is active.

The timed port Automaton $CMA_c$ for communication medium $CM_c$ is defined by:

$CMA_c = (\Sigma, S, S^0, \delta)$
where
$\Sigma = (Origins_c, Destinations_c, \emptyset)$
$S = (Destinations_c \to D^*) \times (Destinations_c \cup \{Nil\})^\omega$
$S^0 = \{(\lambda x.\epsilon, fairlist) | \forall d \in Destinations_c. \#Filter(d, fairlist) = \infty\}$
$\delta = \{((s, d\&l), \theta + \phi, (t, l)) | \phi\&t = s\&distribute(\theta) \wedge (s.d \neq \epsilon \Rightarrow \phi.d \neq \epsilon)\}$

The set of states $S$ is composed of

- a buffer for every destination port,

- a fairness list of destination ports to prevent infinite delay of messages in these buffers.

In each transition, one destination port is removed from the fairness list. If the corresponding buffer is not empty a message to the removed destination port is sent.

To ensure fairness, in the initial fairness list each destination port has to occur infinitely often. $Nil$ denotes, that no output has to occur. Thus only internal actions take place. Moreover, initially all output buffers are empty.

Note that the automaton $CMA_c$ is only weakly pulse-driven.

## 6.9 Distributed Components

Given a distributed component $c \in ID_d$, and assuming given port automata $A_p$ for every part $p \in Parts_c$, together with the automaton $CMA_c$ for the communication medium, a timed port automaton $A_c$ for component $c$ can be composed. As shown in Section 5, there are restrictions on the composability of components. If only weak pulse-driven automata are composed, some kind of blocking may occur. To prevent this, a notion of strongly pulse-drivenness with respect to $(G, P)$ has been developed in the previous section. The definition has also been extended to get well defined composition for infinitely many components.

In Figure 6.1 the composition of distributed component $c$ is shown. It is of the general form for black-box views:

$$CM \otimes (\otimes_{p \in Parts_c} Behavior_c)$$

and for corresponding state-box views:

$$\nu \, (InParts_c \cup OutParts_c) \, : \, CMA_c \otimes (\otimes_{p \in Parts_c} A_c)$$

This composition is not necessarily well defined, due to the fact that all involved automata may be just weakly pulse-driven. However, we give two sufficient conditions for well definedness:

**Theorem 12 (strong pulse-drivenness of communication medium)** *If the communication medium automaton $CMA_c$ is strongly pulse-driven, the composition of $A_p, p \in Parts_c$ and $CMA_c$ is well defined.*

**Proof:** First we observe, that $\otimes_{p \in Parts_c} A_c$ is just a parallel composition of compatible automata without any feedback. According to Theorem 7, we get $(I_j^f, O_j^f) = (\emptyset, \emptyset)$ for every $A_j$ and thus trivially strong pulse-drivenness of $A_j$ with respect to $(I_j^f, O_j^f)$. Thus $A = \otimes_{p \in Parts_c} A_c$ is well defined.

Next, $CMA_c$ and $A$ are compatible automata that are composed by the binary variant of $\otimes$. As $CMA_c$ is strongly pulse-driven, Theorem 4 ensures well definedneess of the overall composition. □

**Theorem 13 (strong pulse-drivenness of parts)** *If the automaton $A_p$ for every part $p$ of a component $c$ is strongly pulse-driven, the composition of $A_p, p \in Parts_c$ and $CMA_c$ is well defined.*

**Proof:** Similarly to the above given proof $A = \otimes_{p \in Parts_c} A_c$ is well defined. As all components are strongly pulse-driven, so is their parallel composition $A$. Again Theorem 4 ensures well definedness of the overall composition. □

Strongly pulse-drivenness of the communication medium can be ensured by restricting the transition relation accordingly. The timed port Automaton $CMAS_c$ for strongly pulse-driven communication medium $CM_c$ can be defined by:

$CMAS_c = (\Sigma, S, S^0, \delta')$
with $\Sigma, S, S^0$ as before and new
$\delta' = \{((\phi\&s, d\&l), \theta + \phi, (s\&distribute(\theta), l)) | (\phi\&s).d \neq \epsilon \Rightarrow \phi.d \neq \epsilon\}$

The input $\theta$ is stored in the next state, whereas the output $\phi$ is taken from the actual state, and therefore has been stored earlier. To keep on progress, for every channel $d$ the output $\phi.d$ is not empty if there are messages in the store $(\phi\&s).d$.

A strongly pulse-driven communication medium imposes a delay of at least one time unit for every message. Such restrictions of the transition relation are allowed, as long as reactivity is preserved, because the above given proofs and theorems do not use the structure of $CMA_c$. Observe, that $\delta' \subseteq \delta$ and due to the refinement calculus for port automata defined in [Rum96]:

$$[CMAS_c] \subseteq [CMA_c]$$

If we want the composition result to be strongly pulse-driven again, we can use the following theorem:

**Theorem 14 (strong pulse-drivenness of composed automaton)** *The composition*

$$CMA_c \otimes (\otimes_{p \in Parts_c} A_c)$$

*is strongly pulse-driven again, if the the automaton $A_p$ for every part $p$ of a component $c$ is strongly pulse-driven, and no message from the environment of $c$ is sent back to this environment by the communication medium CMA:*

$$\forall d \in D. origin_c(d) \in In_c \Rightarrow destination_c(d) \cap Out_c = \emptyset$$

**Proof:** Theorem 13 ensures well definedness of the composition. The above stated condition ensures, that incoming messages (from $In_c$) of a time unit are only sent to the parts of the component. This ensures a strong pulse-drivenness of the communication medium $CMA_c$ with respect to $(In_c, Out_c)$. As all parts are strongly pulse-driven this property is preserved by the composition, meaning $CMA_c \otimes (\otimes_{p \in Parts_c} A_c)$ is strongly pulse-driven with respect to $(In_c, Out_c)$. This is equivalent to strong pulse-drivenness of $CMA_c \otimes (\otimes_{p \in Parts_c} A_c)$. □

From Theorem 14 follows immediately that the use of a strongly pulse-driven automaton $CMAS_c$ always leads to a strongly pulse-driven composition.

# Chapter 7

# Examples

To get a better understanding how the presented system model can be applied, we now give two examples. The first example, a FIFO-Queue, is an example for a typical software system. The second example, an RS-flipflop, is taken from the area of hardware design.

## 7.1 Queue

We model an interactive queue which stores incoming messages of type *Value* and dispenses them on request in a FIFO-order. Because this is a well-known example, we do not specify the black-box behavior of queues here.

The queue has the identifier $qu$, the input port $qu^i$, and the output port $qu^o$. It is implemented by a linked list of *queue elements*. The set of identifiers for the queue elements is given by the infinite set $IDQ \subset ID$. As common in object-oriented systems, message addressing is done via identifiers. Identifiers are sufficient, because we design every component to have exactly one input- and one output port. For every $q \in IDQ$, $q^i$ and $q^o$ denote the input respectively the output port of queue element $q$. $IDQ^i$ and $IDQ^o$ are the corresponding sets of all input- and output ports of queue elements:

$In_q = \{q^i\}, Out_q = \{q^o\}$
$In_{qu} = \{qu^i\}, Out_{qu} = \{qu^o\}$

A queue element contains an ordinary value of set *Value* or the special value *None* to indicate that no ordinary value is contained. If a queue element contains a value, it also contains the identifier of the next queue element (the rest of the queue). Otherwise it uses a special element *Nil* to indicate that no next element exists. Thus a queue element has two attributes:

$val : Value \cup \{None\}$
$next : IDQ \cup \{Nil\}$

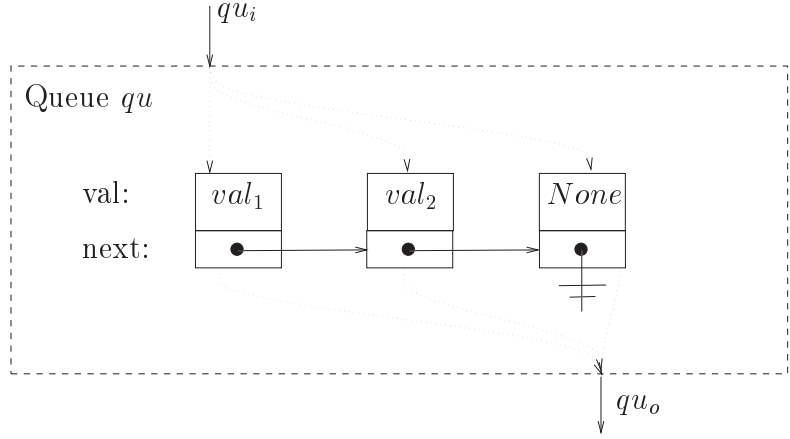

Figure 7.1: Dynamic structure of a queue

At each point of time only a finite subset of queue elements contains values. We therefore may depict the state of a Queue by its finite set of elements containing values. Figure 7.1 depicts a two element queue. The messages of the input port of an element of class Queue are dispatched to its elements. Accordingly the messages from queue elements to the environment are collected and emitted through the output port. The signature of queue elements consists of two kinds of messages, to enque new values at the end of the queue ($enq$) and to deque values at the top of the queue ($deq$). The signature of queue elements is:

$.enq(x : Value)$
$.deq(receiver : ID)$

The dequed value, as well as the identifier of the rest of the queue are sent to the explicitly given receiver with message $deq'd(v, next)$. Besides the above defined signature, every message is in addition augmented with the identifiers of the sender and of the receiver component. The set of messages $D$ is therefore defined by:

$$D = ID \times ID \times \{enq(v), deq(id), deq'd(v, id) | v \in Value, id \in ID\}$$

We have to ensure well definedness of automata composition. Therefore, we decide the communication medium to be strongly pulse-driven. Hence our queue elements only have to be weakly pulse-driven. The communication medium is of type:

$$CM_{qu} : (IDQ^o \cup \{qu^i\} \to [D^*]) \xrightarrow{p} (IDQ^i \cup \{qu^o\} \to [D^*])$$

A dynamic change of the interconnection structure between components is possible via changing attributes containing component identifiers. The dynamic creation of components can be modeled in the following way: Components that did not receive or send a message until a point in time are regarded as not yet created. Components are created if some message is sent to them by another component (the creator). To model this, we give the creator a sufficient list *newIDs* containing the identifiers of all components it can create. Initially the queue element $q_0$, encoding the empty queue, is assumed to be active. The *creational function toCreate* of type

$$toCreate : IDQ \to [IDQ]$$

delivers this list of identifiers creatable by a given component. It defines a partition on $IDQ \backslash \{q_0\}$, because every component can be created exactly by one other component. It also defines a tree with root $q_0$, because no circular creation dependencies are possible. To represent the set of still available identifiers, we use a third attribute *newIDs* : $[IDQ]$ to the state of queue elements. This attribute is implicit, meaning that the programmer has no explicit control over this variable. It is also not intended to be implemented as an attribute, but it is only used for modeling purposes. In concrete implementations using object oriented programming languages, *ID* is implemented by the runtime system of the implementation language.

The state set of a queue element can now be defined as

$$States_q = (Value \cup \{None\}) \times (IDQ \cup \{Nil\}) \times [IDQ]$$

with the initial states

$$State_q^0 = \{(None, Nil, toCreate_q)\}.$$

The transition diagram $\delta_q$ for queue element $q$ determines the state transition relation of a timed port automaton. This simply means every transition takes one unit in time. All messages arriving within one time interval are processed and the response is emitted. As delay is already encoded in the communication medium, the automata for the queue elements need not impose delay.

The transition diagram is constructed in a pattern matching-style, well known from functional programming:

$v, x \in Value; next \in IDQ, r \in ID, newID \in [IDQ]$

| current state | in | out | next state |
|---|---|---|---|
| $(v, next, newID)$ | $enq(x)$ | $next.enq(x)$ | $(v, next, newID)$ |
| $(None, Nil, id \& newID)$ | $enq(x)$ | | $(x, id, newID)$ |
| $(v, next, newID)$ | $deq(r)$ | $r.deq'd(v, next)$ | $(v, next, newID)$ |

The queue elements are not active themselves, they only react to incoming messages by changing state and/or sending new messages. This property allows us to write transition diagrams in such a pattern-matching style. The behavior for combinations of input messages and receiving states where no entry in the above table occurs is underspecified. This means that if such messages are received in the corresponding states, chaotic behavior is allowed.

At last, the two functions $origin_{qu}$ and $destination_{qu}$ are defined to model how messages are delivered by the communication medium:

$$\forall d, id; \forall qsnd, qrec \in IDQ; osnd, orec \in ID\backslash IDQ.$$
$$origin_{qu}(qsnd, id, d) = qsnd_o$$
$$origin_{qu}(osnd, id, d) = qu_i$$
$$destination_{qu}(id, qrec, d) = \{qrec_i\}$$
$$destination_{qu}(id, orec, d) = \{qu_o\}$$

Note that no message duplication and no message loss is possible, as this is usual in many software systems.

We are now able to compose all the queue elements together with the communication medium to the complete queue automaton $A_{qu}$. We also hide the input- and output ports of the queue elements:

$$A_{qu} = \nu \, (IDQ^i \cup IDQ^o) \, : \, CM_{qu} \otimes (\otimes_{q \in IDQ} A_q)$$

The composition is well defined due to Theorem 12. We get an infinite product of states $A_{qu}$. However, due to this modeling, at every unit of time just a finite subset of components can be active.

To reason about the overall behavior of the complete queue, one can and should use an abstraction of this state set and define an abstract timed port automaton that can be refined into $A_{qu}$.

## 7.2 RS-Flip-Flop

As a second example a standard hardware component, an RS-flip-flop, is modeled. As depicted in Figure 7.2 it consists of two nor-gates.

The nor-gates are statically connected through channels ("wires"). Therefore, in 7.2 we need not show the communication medium, but show only the static connection of the nor-gates. This static connection is nevertheless realized by a communication medium, which is however very simple (see below).

The channels do not have any delay. Delay of exactly one time unit is introduced by the nor-gates. Note that each output channel $o_j$ branches into an output channel and into a feedback channel. Hardware usually has a time synchronous behavior. Therefore, in our model in each time unit exactly one message (here simply $O$ or $L$) occurs.

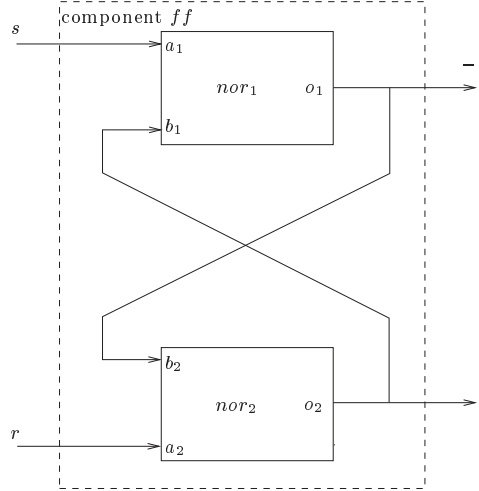

Figure 7.2: RS-flip-flop example

If *ff* is the identifier of the flip-flop, let $nor_1$ and $nor_2$ denote the nor-gates used to construct the flip-flop, i.e. we have $Parts_{ff} = \{nor_1, nor_2\}$. The signature of the three components can is defined as follows:

$$In_{ff} = \{r, s\} \qquad Out_{ff} = \{q, \bar{q}\}$$
$$In_{nor_j} = \{a_j, b_j\} \qquad Out_{nor_j} = \{o_j\}$$

Messages between these hardware components are nothing else than logical $L$ and $O$. However, in the system model all messages are attributed with sender and receiver ports. Due to direct wiring it is enough to attach the port name of the sender:

$$D = \{O, L\} \times P$$

As the receiver doesn't use the additional information that is given by the sender portname, it is not necessary to implement this information. This allows the usual hardware implementation where data of form $\{O, L\}$ is transmitted via wires.

A nor-gate doesn't have any state besides that it delays the output of a message for one time unit. Thus a nor gate is strongly pulse-driven. This modeling is adequate, because the duration of a time unit may be chosen such that every component shows delay [Fuc94]. A *nor* gate therefore is modeled with the state space

$$States_{nor_j} = \{O, L\}$$

and (as usual in hardware) an unknown respectively free selectable initial state, i.e. $State^0_{nor_j} = States_{nor_j}$. The transition relation is defined as follows:

$$\delta_{nor_j} = \{(s, (a_j \mapsto (a,x); b_j \mapsto (b,y); o_j \mapsto (s, o_j)), t)|\\ (a,x), (b,y) \in D \wedge t = nor(a,b)\}$$

$\delta_{nor_j}$ is not reactive as specified above, because it doesn't define any transition that allows input of more than one message a time unit. Due to our assumption on the environment, which required that exactly one message will occur on every line in each time interval, is however sufficient. To get a reactive automaton $\delta_{nor_j}$, we extend the automaton by adding auxiliary transitions for the other input possibilities. These transitions may for instance lead to a state in which arbitrary (chaotic) behavior is possible (see [GKRB96]). While chaotic behavior is adequate in software systems, for the modeling of hardware systems a less liberal behavior in these cases might be adequate. The reason is that sending a gate more than one message within one interval of time usually will not lead to arbitrary behavior in the future, but it may only influence the output of the gate for instance in the subsequent time interval.

We will now specify the communication medium of the flip-flop. The communication structure is given by functions $origin_{ff}$ and $destination_{ff}$:

$$\forall p. origin_{ff}((b,p)) = p\\ origin_{ff}(m) = s \Rightarrow destination_{ff}(m) = \{a_1\}\\ origin_{ff}(m) = r \Rightarrow destination_{ff}(m) = \{a_2\}\\ origin_{ff}(m) = o_1 \Rightarrow destination_{ff}(m) = \{\bar{q}, b_2\}\\ origin_{ff}(m) = o_2 \Rightarrow destination_{ff}(m) = \{q, b_1\}$$

The communication medium doesn't delay any messages. It therefore has just one state and it is only weakly pulse-driven. However, the composition of automata is well defined because the nor-gates are strongly pulse-driven. The composed automaton is also strongly pulse-driven due to Theorem 14.

We get the following composed automaton $A_{ff}$, which is (up to isomorphic renaming) depicted in Figure 7.3. In the picture, irregular inputs are omitted for simplicity. As always in strongly pulse-driven automata, the output of a transition is fully determined by its source state and therefore left out in the transitions.

A state of a system is called *stable*, if it does not change without further input from the environment. From this automaton it is very easy to conclude how many time units it takes to get a stable state for a given input. One can also see that for input $O, O$ no stable state is reached.

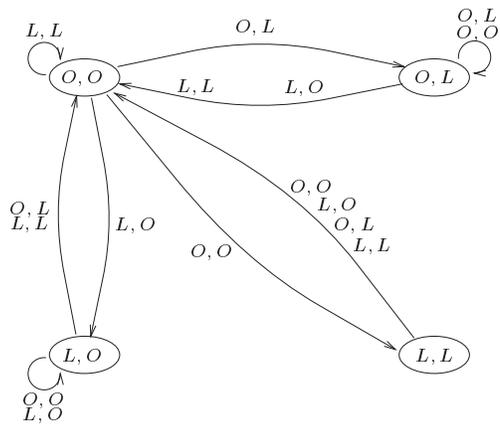

Figure 7.3: Port automaton for the RS-flip-flop

# Chapter 8

# Conclusion

In this paper, we integrated the SysLab system model ([RKB95], [KRB96]) with port automata [GR95]. To achieve this, port automata have been enhanced by an infinite composition operator.

The SysLab system model has been enhanced by a state based modeling technique, allowing the connection between behavior of components and the state of components. Two different approaches allow to compose state based descriptions of components together with the communication medium, yielding a new state based description of the composed component.

Minor enhancements to the system model are the generalized message passing mechanism, that comprises channel based message sending as well as broadcasting, and the introduction of types for channels and state spaces.

This paper defines an additional layer to the original system model. Its purpose is to provide a rigorous mathematical basis to define the semantics of the different SysLab description techniques that are in development.

## Thank

For a careful reading of draft versions of the paper and stimulating discussions we thank Barbara Paech, Manfred Broy, Peter Scholz and Ketil Stølen.